\documentclass[a4,12pt,eclepsf]{article}
\usepackage[dvips]{graphicx,color}
\usepackage{amsmath}
\usepackage{amssymb}

\begin{document}

\renewcommand{\theequation}
{\thesection.\arabic{equation}}
\thispagestyle{empty}
\vspace*{20mm} 
\begin{center}
{\LARGE {\bf Glueball mass spectra for supergravity duals of}} \\
 \qquad \\
{\LARGE {\bf noncommutative gauge theories}} \\
 \qquad \\

\vspace*{10mm}

\renewcommand{\thefootnote}{\fnsymbol{footnote}}

{\large Tadahito NAKAJIMA, \footnotemark[1] 
 \, Kenji SUZUKI \footnotemark[2] and 
 Hidenori TAKAHASHI \footnotemark[3]} \\ 

\vspace*{10mm}

$\footnotemark[1]$ {\it Research Institute of Science and Technology, Nihon 
University, Tokyo 101-8308, Japan} \\
$\footnotemark[2]$
{\it Department of Physics, Ochanomizu University, Tokyo 112-8610, Japan} \\
$\footnotemark[3]$
{\it Laboratory of Physics, College of Science and Technology, 
Nihon University, Chiba 274-8501, Japan} \\

\vspace*{20mm}

\footnotetext[1]
{E-mail: nakajima@phys.cst.nihon-u.ac.jp}
\footnotetext[2]
{E-mail: ksuzuki@cc.ocha.ac.jp}
\footnotetext[3]
{E-mail: htaka@phys.ge.cst.nihon-u.ac.jp} 

\vspace*{5mm}

\quad 

\vspace*{5mm}

{\bf Abstract} \\

\end{center}

We derive the glueball masses in noncommutative super Yang--Mills theories 
in four dimensions via the dual supergravity description. The spectrum of 
glueball masses is discrete due to the noncommutativity and the glueball 
masses are proportional to the noncommutativity parameter with dimension of 
length. The mass spectrum in the WKB approximation closely agrees with the 
mass spectrum in finite temperature Yang--Mills theory.

\vspace*{15mm}

\clearpage

\setcounter{section}{0}
\section{Introduction}
\setcounter{page}{1}
\setcounter{equation}{0}

In recent years, it has been possible to investigate aspects of the large $N$ 
noncommutative gauge theories in the spirit of the AdS/CFT 
\cite{malda, gub_kle_pol, wit} correspondence. Noncommutative 
gauge theories arise as a certain low-energy limit of string theory in 
Neveu-Schwarz-Neveu-Schwarz (NS-NS) B-field background \cite{sei_wit}. 
Supergravity duals of large N noncommutative gauge theories with maximal 
supersymmetry have been constructed as the decoupling limits of D-brane solutions with NS-NS B fields \cite{has_itz, malda_russ}. 
Noncommutative gauge theories 
are intriguing dynamical systems which exhibit rich features such as gauge 
invariance, nonlocality and UV/IR mixing. These supergravity solutions have 
been used to investigate qualitative aspects of nonperturbative gauge theories 
\cite{Ali_Oz_Jab, Gross_Hashimoto_Itzhaki, Cai_Ohta1, Cai_Ohta2, Cai_Ohta3, 
Berman}. Since noncommutativity introduces a new physical 
scale to the theories, it modifies the Wilson loop behavior. If 
noncommutativity effects are large, then they exhibit area law 
\cite{dhar_kita, lee_sin, taka_naka-suzu}. Supergravity duals of 
noncommutative gauge theories with less than maximal supersymmetry have also 
been constructed \cite{Bru_Gom_Mat_Ram, mat_pon_tal}. 
The behavior of the Wilson loops in ${\cal N}=1$ NCSYM theory has been 
investigated from a deformation of the Maldacena--N\`u\~nez solution, which is 
proposed as supergravity duals of ${\cal N}=1$ NCSYM theory. 
The quark-antiquark potential via the Wilson loop gives a same behavior 
as ordinary ${\cal N}=1$ super Yang--Mills theory in the IR region, 
although the UV physics give a different behavior. The 
$\beta$-function in the ${\cal N}=1$ NCSYM theory has also been computed and 
the $\beta$-function of the NCSYM theory coincides with the ordinary one 
\cite{mat_pon_tal}. 

It is well known that noncommutative gauge theories have no local gauge 
invariant operators. Nevertheless there are non-local gauge invariant 
operators which are the Fourier transform of local operators attached to 
open Wilson lines \cite{Liu, Meh_Wis}. It seems to indicate that the 
supergravity fields act as sources of such kind of gauge invariant 
operators \cite{Liu_Mic, Oka_Oog, Das}. The fact that supergravity 
fields do not depend on the noncommutative coordinates makes it easier to 
obtain the gravity fields that are dual to such kind of gauge theory operators.

Supergravity solutions have been used to study qualitative 
aspects of non-perturbative gauge theories not only quark confinement, 
but also chiral symmetry breaking, renormalization group flow, binding energy 
of the baryon and glueball mass spectrum.  A discrete glueball spectrum with 
a finite gap can be derived by compactification of the dual supergravity 
models. The radius of the compactifying circle provides the ultraviolet cut-off 
scale and the glueball masses are measured in unit of the compactification 
radius. The ratio of the glueball masses is a fairly good quantitative 
agreement with lattice data \cite{csa_oog_oz_ter, Koc_Jev_Mih_Nun, Zys, mina}. 

In this paper, we study some nonperturbative aspects of noncommutative 
Yang--Mills (NCYM) theories by focusing on evaluating 
mass spectra of the glueballs. Since NCYM theories have an intrinsic 
physical scale, there is a possibility that the physical scale reflects 
the discrete mass spectra without any compactifications.
The paper is organized as follows. In Section 2, we evaluate $0^{++}$ glueball 
masses in noncommutative super Yang--Mills (NCSYM) theory in four dimensions 
by solving the wave equations for dilaton in the dual supergravity background. 
The mass eigenvalues can be determined approximately via the 
Wentzel-Kramers-Brillion (WKB) analysis. 
$0^{++}$ glueball masses in NCSYM theory in a constant self-dual of gauge 
field background are also evaluated using the dual supergravity description. 
In section 3, we evaluate $1^{--}$ glueball masses in NCSYM theory in four 
dimensions by solving the wave equations for antisymmetric tensor field in 
the dual supergravity background. All the results are compared with the 
glueballs masses in finite temperature Yang--Mills theories from supergravity 
computation and lattice computations. Section 4 is devoted to conclusions and 
discussions.

%
%

%
\section{The $0^{++}$ glueball masses in noncommutative gauge theory}
\subsection{Glueball masses in NCSYM theory}
\setcounter{equation}{0}

We begin with the D3 brane solution in a NS-NS B-field 
background in the near horizon limit \cite{liu_tse, lee_sin}: 
\begin{eqnarray}
ds^2 = \alpha'R^2 
[u^2 \{-d x^2_0 + d x^2_1 + \hat{h} \ (d x^2_2 + d x^2_3) \} 
    + ( \frac{d u^2}{u^2} + d \Omega^2_5 )] \;, 
\end{eqnarray}
%
where
\begin{eqnarray}
& & \hat{h}^{-1} = 1 + a^4u^4 \;. 
\end{eqnarray}
%
Here we assume that the NS-NS B field has the non-vanishing component of 
$B_{23}$. In order to obtain NCSYM theory we should take the B-field to 
infinity in the near horizon limit as $B \alpha'= fixed$. The 
noncommutativity parameter $a$ is related to the rescaling B-field 
$\widetilde{B}_{23}$ as 
$\widetilde{B}_{23}=\frac{\alpha'R^2}{1 + a^4u^4}$.

The $0^{++}$ glueball masses can be derived from the 2-point function
of the dimension 4 scalar operators ${\cal O}_{4} = tr F^2$. The scalar 
operators ${\cal O}_{4}$ couples to the real part of a complex massless scalar 
field that consists of the dilaton and the Ramond-Ramond (R-R) scalar field. 
When we evaluate the $0^{++}$ glueball masses, we have to solve the classical
equation of motion of the massless dilaton in the supergravity background 
\cite{csa_oog_oz_ter, mina}. Consider the wave equation for the dilaton: 
\begin{eqnarray}
\partial_\mu \,\left\{\, e^{-2\phi} \sqrt{g}g^{\mu\nu}\partial_\nu \phi
\,\right\}\, = 0 \,.
\end{eqnarray}
Under the metric of (2.1) the dilaton equation (2.3) is given by: 
\begin{eqnarray}
\partial_u \,[\, u^5 \partial_{u} \rho\,]\, 
+ u(k_0^2 - k_1^2 )\rho 
- (1 + a^4u^4)u (k_2^2 + k_3^2)\rho = 0 \, .
\end{eqnarray}
%
In deriving this equation, we assume that the dilaton $\phi$ has the plane 
wave form $\phi = e^{ik \cdot x}\rho(u)$. The glueball mass $M^2$ is equal 
to $-k^2$. In order to take in the effects of the noncommutativity to the 
wave equation, we choose a particular momentum 
$k^{\mu}=(\frac{M}{\sqrt{1-\beta^2}}, 0, \frac{\beta M}{\sqrt{1-\beta^2}}, 0)$ 
that is given by {\it the Lorentz boost} of the rest frame momentum 
$k^{\mu}=(M, 0, 0, 0)$. In other words, we consider the dilaton equation in 
the ``moving" frame with the velocity $\beta$ in unit of the light velocity
\cite{dhar_kita, Ali_Oz_Jab, mat_pon_tal}. 
Then the equation (2.4) becomes 
\begin{eqnarray}
\partial_u \,[\, u^5 \partial_u \rho\,]\, 
+ \frac{M^2}{1-\beta^2}u [1-\beta^2 (1 + a^4u^4 )] \rho = 0 \;.
\end{eqnarray}
%
When we change the variable to $y=u^2$, the equation takes the form 
\begin{eqnarray}
\partial_y \,[\, y^3 \partial_y \rho\,]\, 
+ \frac{M^2}{4(1-\beta^2)}[1-\beta^2 (1 + a^4y^2 )] \rho = 0 \;. 
\end{eqnarray}
%
Since the differential equation (2.6) has singularities at $y=0$ and 
$y \rightarrow \infty$, 
we rewrite the equation (2.6) by using a new variable $a^{2}y=e^{z}$. Then we 
have 
\begin{eqnarray}
\partial_z \,[\, e^{2z} \partial_z \rho \,]\, 
+ \frac{M^2a^2}{4}e^{z}[1 - \gamma e^{2z}] \rho = 0 \;, 
\end{eqnarray}
%
with $\gamma \equiv \frac{\beta^2}{1-\beta^2}$. 
For a definition of the new function $\rho = e^{-z}\psi$ we can obtain the 
Schr\"{o}dinger-type equation as 
\begin{eqnarray}
\psi'' + V\psi = 0 \;,
\end{eqnarray}
%
where $'$ denotes the differentiation with respect to the variable $z$. 
The explicit form of the potential for the Schr\"{o}dinger equation (2.8) is 
given by 
\begin{eqnarray}
V = - \frac{1}{4}M^2 a^2 \gamma e^{-z} (e^z-a^2 \lambda_{+})
(e^z-a^2 \lambda_{-}) \, \;,
\end{eqnarray}
%
where 
\begin{eqnarray}
\lambda_{\pm} = -\frac{2}{M^2 a^4 \gamma} \left\{ 
1 \pm \sqrt{1 + \frac{M^4 a^4 \gamma}{4}} 
\right\} \;. 
\end{eqnarray}
%
This potential has the turning points at $z=\ln (a^{2}\lambda_{-})$. 
We shall evaluate the mass spectrum within the semiclassical WKB 
approximation. The WKB approximation for this potential gives 
\begin{eqnarray}
\left( n + \frac{1}{2} \right)\pi \!\!&=&\!\!
\int^{\ln \lambda_{-}}_{-\infty} dz \sqrt{V} \nonumber \\
\!\!&=&\!\! \sqrt{\frac{1}{4}M^2 a^4 \gamma}
\int^{\lambda_{-}}_{0} dy 
\sqrt{\frac{(y-\lambda_{+})(\lambda_{-}-y)}{y^3}} \;, 
\end{eqnarray}
%
where $n$ denotes the integer. 
By substituting the variable $y=\lambda_{-}t$ 
into the last line of Eq. (2.11), we can rewrite the WKB integral as 
\begin{eqnarray}
\left( n + \frac{1}{2} \right)\pi \!\!&=&\!\!
\sqrt{\frac{1}{4}M^2 a^4 \gamma \,(-\lambda_{+}) } 
\int^{1}_{0} dt \; t^{-3/2}(1-t)^{1/2} \nonumber \\ 
& & \qquad \times \left\{ 
1 - \frac{1}{2}\frac{\lambda_{-}}{\lambda_{+}}t 
- \sum^{\infty}_{m=2}\frac{(2m-3)!!}{2^n} 
\left( \frac{\lambda_{-}}{\lambda_{+}}t\right)^m \right\} \;. 
\end{eqnarray}
%
In deriving Eq. (2.12), we have expanded the expression 
$\sqrt{1-\frac{\lambda_{-}}{\lambda_{+}}t}$ in the Taylor's 
series by taking account of the fact that $0 < 
-\frac{\lambda_{-}}{\lambda_{+}}t < 1$. 
The magnitude of the parameter $\frac{\lambda_{-}}{\lambda_{+}}$ takes values 
smaller than $0.1$ for the noncommutativity parameter $a \sim M^{-1}$ and the 
velocity $\beta<0.8$. Even though the noncommutativity parameter $a$ takes 
large value such as $a \sim 10^{4}M^{-1}$, the parameter 
$\frac{\lambda_{-}}{\lambda_{+}}$ takes such a sufficiently small values in the 
low velocity $\beta<0.014$. Hereafter we restrict our computation within 
the low velocity region where the perturbative analysis is appropriate. 
The right hand side of the expression (2.12) is given as the function 
of the dimensionless quantity $(Ma)^{4}$. 
The glueball masses are obtained by solving the WKB approximation 
(2.12) with respect to the quantity $(Ma)^{4}$ after carrying out the 
integration. 
Up to the leading order in the parameter $\frac{\lambda_{-}}{\lambda_{+}}$ 
we obtain 
\begin{eqnarray}
(Ma)^{4}=\frac{(2n+1)^{2}(2n-1)(2n+3)}{\gamma} \;.
\end{eqnarray}
%
Here we have utilized a regularization based on the analytic continuation for 
Euler's integral of the first kind: 
$\int^{1}_{0} dt \,t^{-3/2}(1-t)^{1/2} \equiv \lim_{p \rightarrow -\frac{1}{2}}
B(p,\;\frac{3}{2})=-\pi$, where $B(p,\;q)$ denotes the Euler's beta function. 
The mass spectrum for $0^{++}$ glueball is given by  
\begin{eqnarray}
M^{0^{++}}_{(L)}=\frac{1}{a}\sqrt[4]{\frac{(2n+1)^{2}(2n-1)(2n+3)}{\gamma}} \;,
\end{eqnarray}
%
The glueball masses (2.14) takes real numbers for positive integer $n =1,2,3, 
\ldots.$ Notice that the glueball masses are proportional to the inverse of 
the noncommutativity parameter $a$. When we take the commutative limit 
$a \rightarrow 0$, then the masses do not take the discrete values. 

Up to the subleading order in the parameter $\frac{\lambda_{-}}{\lambda_{+}}$ 
of the WKB approximation leads the mass spectrum for $0^{++}$ glueball: 
\begin{eqnarray}
M^{0^{++}}_{(L+SL)}=\frac{1}{a} \sqrt[4]{ \frac{4}{81} 
\frac{f_{+}(n)}{\gamma}} \;, 
\end{eqnarray}
%
where $f_{\pm}(n)$ denotes some function of the positive integer $n$ whose 
explicit form is given by 
\begin{eqnarray}
f_{\pm}(n) \!&=&\! 512n^{4}+1024n^{3}+96n^{2}-416n+8 \nonumber \\
\!&\pm&\! 8(16n^{2}+16n-11)\sqrt{(2n+1)^{2}(4n^{2}+4n-2)} \;.
\end{eqnarray}
%
The glueball masses $M^{0^{++}}_{(L+SL)}$ take the positive real 
eigenvalues, while the other choice of glueball masses 
$\widetilde{M}^{0^{++}}_{(L+SL)} 
\equiv \frac{1}{a} \sqrt[4]{\frac{4}{81} \frac{f_{-}(n)}{\gamma}}$ 
take complex eigenvalues. 
The states for $0^{++}$ glueball with the masses 
$\widetilde{M}^{0^{++}}_{(L+SL)}$ are unstable. As will be seen later, 
however, these unstable states are avoidable 
by virtue of introducing a constant self-dual gauge field background.

As was expected that the glueball masses are given in units of the 
noncommutativity parameter $a$. The noncommutativity parameter 
in the NCYM theory plays a similar role to a compactification radius in the 
Yang--Mills theory compactified on a circle, or the temperature in the 
Yang--Mills theories at finite temperature, where the temperature is 
proportional to the inverse of the compactification radius. 
\cite{csa_oog_oz_ter, mina, csa_rus_sfe_ter}. 
We should notice that the spatial noncommutativity make it possible to 
obtain discrete mass spectrum in the Yang--Mills theory without any 
compactifications.

Although the expression (2.15) is a bit complicated, there is little 
difference between the ratios of the $0^{++}$ glueball masses up to the 
leading order and the subleading order. The ratios of the $0^{++}$ glueball 
masses $M^{0^{++}}_{(L+SL)}$ obtained by solving the dilaton wave equation 
in the WKB approximation are listed in Table 1. 
\begin{table}[htbp]
\begin{center}
\begin{tabular}{|c|c|c|}
\hline
state & WKB(up to leading) & WKB(up to subleading) \\
\hline 
$0^{++}$ & 1\,(input) & 1\,(input) \\
$0^{++*}$ & 1.85 & 1.89 \\
$0^{++**}$ & 2.64 & 2.73 \\
$0^{++***}$ & 3.43 & 3.56 \\
\hline
\end{tabular}
\caption{Masses of the $0^{++}$ glueball in NCYM${}_{4}$.}
\end{center}
\end{table} 

The glueball masses also depend on the boost parameter $\gamma$, 
besides the noncommutativity parameter $a$. 
When we evaluate the glueball masses $M$ in the rest flame with the boost 
parameter $\gamma=0$, then the glueball can not take the discrete 
mass spectrum with a finite gap. 
This is consistent with the fact that the effects of the noncommutativity 
are taken in by the moving frame. 
The ratio of the masses does not depend not only on the noncommutativity 
parameter $a$, but also on the boost parameter $\gamma$, which is a 
dimensionless parameter. This fact is not an 
accident. We can regard the WKB integral (2.12) as an algebraic equation for 
the variable $M^{4}a^{4}\gamma$. If we can solve the algebraic equation, then 
we have the variable $M^{4}a^{4}\gamma$ as a function of the integer $n$. 
therefore, the glueball masses are also a function of the integer $F(n)$ 
as $\displaystyle{M=a^{-1}\gamma^{-1/4}F(n)}$. Although the masses depend 
on the noncommutativity parameter $a$ and the boost parameter $\gamma$, the 
ratios of the masses are independent of both parameters.
Comparison of $0^{++}$ glueball masses in finite temperature Yang--Mills 
theory in four dimensions from supergravity, besides of the lattice QCD results in four dimensions is shown in Table 2. 
\begin{table}[htbp]
\begin{center}
\begin{tabular}{|c|c|c|c|}
\hline
state & $NCYM_{4}$\,(WKB) & $QCD_{4}$\,(WKB) \cite{mina} 
& $QCD_{4}$\,(Lattice) \cite{tep, mor_pea}\\
\hline 
$0^{++}$ & 1\,(input) & 1\,(input) & 1\,(input) \\
$0^{++*}$ & 1.89 & 1.62 & 1.75 ($\pm$0.17) \\
$0^{++**}$ & 2.73 & 2.24 & - \\
$0^{++***}$ & 3.56 & 2.82 & - \\
\hline
\end{tabular}
\caption{Masses of the $0^{++}$ glueball from supergravity 
and lattice QCD.}
\end{center}
\end{table} 
From Table 2 we find that the difference between the supergravity or 
lattice results in the QCD and the supergravity ones in the NCYM theory is 
small. 

%
%

%
\subsection{Glueball masses in NCSYM theory in a constant self-dual 
background}  

In this subsection we evaluate the glueball masses in the four dimensional 
NCSYM theory in a constant self-dual gauge field background using the 
dual supergravity description. Its supergravity dual is known as a limit of superposition of D3-brane and D(-1)-brane (D-instanton) backgrounds 
\cite{liu_tse, lee_sin}. 
The metric for the supergravity solution in the near horizon limit is: 
\begin{eqnarray}
ds^2 = \alpha'R^2 \left(\; 1+\frac{q}{R^4u^4} \;\right)^{1/2} 
[u^2 \{-d x^2_0 + d x^2_1 + \hat{h} \ (d x^2_2 + d x^2_3) \} 
    + ( \frac{d u^2}{u^2} + d \Omega^2_5 )] \;, 
\end{eqnarray}
%
where 
\begin{eqnarray*}
\hat{h} = \frac{1}{1+Ha^{4}u^{4}} \;, 
\end{eqnarray*}
%
with $\displaystyle H=1+\frac{q}{R^{4}u^{4}}$. Here $q$ denotes the 
{\it D}-instanton density. 

Under the metric of (2.17) we obtain the wave equation for the dilaton 
$\phi = e^{ik \cdot x}\rho(u)$  
\begin{eqnarray}
\partial_u \,[\, u^5 \partial_u \rho\,]\, 
+ \frac{M^2}{1+\beta^2}u [1+\beta^2 (1 + H a^4u^4 )] \rho = 0 \;.
\end{eqnarray}
with a particular momentum $k^{\mu}=(\frac{M}{\sqrt{1+\beta^2}}, 0, 
\frac{\beta M}{\sqrt{1+\beta^2}}, 0)$. Changing the variable to $z=2\ln (au)$ 
we have 
\begin{eqnarray}
\partial_z \,[\, e^{2z} \partial_z \rho \,]\, 
+ \frac{M^2 a^2}{4} e^{z}[1 + \gamma H e^{2z}] \rho = 0 \;, 
\end{eqnarray}
%
where $\gamma = \frac{\beta^2}{1+\beta^2}$. For a 
redefinition of a function $\rho = e^{-z}\psi$ we can obtain the 
Schr\"odinger-type equation as 
\begin{eqnarray}
\psi'' + V\psi = 0 \;.
\end{eqnarray}
%
The potential $V$ takes the form:  
\begin{eqnarray}
V = -\frac{1}{4}M^2 a^2 \gamma e^{-z} (e^z-a^2 \kappa_{+}) (e^z-a^2 \kappa_{-}) 
\, \;,
\end{eqnarray}
%
where 
\begin{eqnarray}
\kappa_{\pm} = -\frac{2}{M^2 a^4 \gamma} \left\{ 
1 \pm \sqrt{1 + \frac{M^4 a^4 \gamma}{4}(1-\frac{a^4 q}{R^4}\gamma)} 
\right\} \;. 
\end{eqnarray}
%
There is also a turning point at $z=\ln (a^{2} \kappa_{-})$. If the parameter 
$\gamma$ satisfies the condition: 
\begin{eqnarray}
0 < \gamma < \frac{R^4}{2a^4 q} 
\left\{ \sqrt{1+\frac{16q}{M^4R^4}} -1 \right\} 
\;, 
\end{eqnarray}
%
both of the parameters $\kappa_{\pm}$ become real (and positive) 
numbers. The condition (2.23) shows that the boost parameter $\gamma$ is 
restricted within a certain range when $\frac{R^4}{2a^4 q} 
\left\{ \sqrt{1+\frac{16q}{M^4R^4}} -1 \right\}$ is smaller than $1$. 
%
%
The WKB approximation for this potential gives 
\begin{eqnarray}
\left( n + \frac{1}{2} \right)\pi \!\!&=&\!\!
\sqrt{\frac{1}{4}M^2 a^4 \gamma}
\int^{\kappa_{-}}_{0} dy 
\sqrt{\frac{(y-\kappa_{+})(\kappa_{-}-y)}{y^3}} \;, 
\end{eqnarray}
%
where $n$ denotes the integer. Here we have rewritten the WKB integral (2.24) 
by using the variable $y=a^{-2}e^{z}$. By substituting the variable 
$y=\kappa_{-}t$ into Eq. (2.24), we can rewrite the WKB integral as 
\begin{eqnarray}
\left( n + \frac{1}{2} \right)\pi 
\!\!&=&\!\! \sqrt{\frac{1}{4}M^2 a^4 \gamma \kappa_{+}} 
\int^{1}_{0} dt \; t^{-3/2}(1-t)^{1/2} \nonumber \\ 
& & \qquad \times \left\{ 
1 - \frac{1}{2}\frac{\kappa_{-}}{\kappa_{+}}t 
- \sum^{\infty}_{m=2}\frac{(2m-3)!!}{2^n} 
\left( \frac{\kappa_{-}}{\kappa_{+}}t\right)^m \right\} \;. 
\end{eqnarray}
%
In deriving Eq. (2.25), we have expanded the expression 
$\sqrt{1-\frac{\kappa_{-}}{\kappa_{+}}t}$ in the Taylor's series by 
taking account of the fact that $0 < -\frac{\kappa_{-}}{\kappa_{+}}t < 1$.

The glueball masses are obtained by solving the Eq. (2.25) for $M$. Up to the 
leading order in the parameter $\frac{\kappa_{-}}{\kappa_{+}}$ 
the $0^{++}$ glueball masses are given by 
\begin{eqnarray}
M^{0^{++}}_{(L)}=\frac{1}{a} \sqrt[4]{ \frac{(2n+1)^{2}(2n-1)(2n+3)}
{\gamma \left( 1 -a^{4}\gamma \frac{q}{R^{4}} \right)} } \;. 
\end{eqnarray}
%
When the boost parameter $\gamma$ satisfied the condition 
\begin{eqnarray}
0 < \gamma < \frac{R^{4}}{a^{4}q} \;, 
\end{eqnarray}
%
the mass spectrum (2.26) takes real numbers for positive integer $n =1,2,3, 
\ldots.$ The condition (2.27) is stronger than the condition (2.23). The 
boost parameter $\gamma$ is restricted by the instanton density $q$. 

We next evaluate the glueball masses up to the subleading order in 
the parameter $\frac{\kappa_{-}}{\kappa_{+}}$. The mass spectrum is given by 
\begin{eqnarray}
M^{0^{++}}_{(L+SL)}=\frac{1}{a}\sqrt[4]{
\frac{2f_{+}(n)}{9\gamma \left( 1 -a^{4}\gamma \frac{q}{R^{4}} \right)}} \;, 
\end{eqnarray}
%
with the condition (2.27) and 
\begin{eqnarray}
\widetilde{M}^{0^{++}}_{(L+SL)}=\frac{1}{a}\sqrt[4]{
\frac{2f_{-}(n)}{9\gamma \left(1-a^{4}\gamma \frac{q}{R^{4}} \right)}} \;, 
\end{eqnarray}
%
with the condition $\displaystyle{\frac{R^{4}}{a^{4}q} < \gamma \;}$. 
%
%
Here $f_{\pm}(n)$ is the same function as (2.16). When we take the limit 
$q \rightarrow 0$, then the glueball masses (2.29) take 
the complex values and the states for $0^{++}$ glueball 
with the masses (2.29) becomes unstable. By virtue of the instanton 
effects, the states for $0^{++}$ glueball with the masses (2.29) becomes 
stable. 
Although the instanton effects changes the $0^{++}$ glueball masses, they do 
not affect the ratio of the $0^{++}$ glueball masses. The ratios of the 
$0^{++}$ glueball masses $M^{0^{++}}_{(L+SL)}$ and $\widetilde{M}^{0^{++}}_{(L+SL)}$ 
in NCYM theory in a constant self-dual background are listed in Table 3. 
\begin{table}[htbp]
\begin{center}
\begin{tabular}{|c|c|c|}
\hline
state & $M$ (up to subleading) & $\widetilde{M}$ (up to subleading) \\
\hline 
$0^{++}$ & 1\,(input) & 1\,(input) \\
$0^{++*}$ & 1.89  &  0.75 \\
$0^{++**}$ & 2.73 &  0.63 \\
$0^{++***}$ & 3.56 &  0.56   \\
\hline
\end{tabular}
\caption{Masses of the $0^{++}$ glueball in NCYM in a constant self 
dual background.}
\end{center}
\end{table} 
The ratios of the glueball masses $\widetilde{M}^{0^{++}}_{(L+SL)}$ is different 
from that of the glueball masses $M^{0^{++}}_{(L+SL)}$. 

%
%
\section{The $1^{--}$ glueball masses in noncommutative gauge theory}
\setcounter{equation}{0}

We next evaluate the $1^{--}$ glueball masses in NCSYM theory in four 
dimensions using the dual supergravity description. 
The $1^{-+}$ and $1^{--}$ glueball masses can be derived from the 2-point 
function of the dimension 6 two-form operators ${\cal O}_{6} = d^{abc} 
F_{\mu\rho}{}^{a}F^{\rho\sigma}{}^{b}F_{\sigma\nu}{}^{c}$, where $d^{abc}$ 
is the symmetric structure constant. 
The two-form operator ${\cal O}_{6}$ couples to the real part of a 
complex-valued antisymmetric tensor field $A_{\mu\nu}$ field which consists 
of the NS-NS and R-R two-forms fields. The operator contains $1^{-+}$ and 
$1^{--}$ components, which correspond to the fields $A_{0i}$ and $A_{ij}$, 
where  $i, j = 1, 2, 3$ correspond to the three coordinates of 
$\mbox{\boldmath $R$}^{3}$. When we evaluate the $1^{-+}$ and $1^{--}$ 
glueball masses, we have to solve the classical equation of motion of 
the massless antisymmetric tensor field $A_{\mu\nu}$ in the supergravity 
background \cite{csa_oog_oz_ter}.  

Consider the wave equation for the complex-valued antisymmetric tensor 
field $A_{\mu\nu}$: 
\begin{eqnarray}
\partial_\mu \,\left\{\, \sqrt{g} \partial_{[\,\kappa}A_{\rho\sigma\,]}
g^{\mu\kappa}g^{\rho\nu}g^{\sigma\lambda} \,\right\}\, = 0 \; ,
\end{eqnarray}
%
where the square brackets $[\;]$ denotes antisymmetrization with the indices. 
Assuming the simplest ansatz we take only one component of the fluctuation. 
Under this assumption, the $1^{-+}$ or 
$1^{--}$ glueball mass spectrum depends on the components of the fluctuation, 
since the metric (2.1) is anisotropic in $\mbox{\boldmath $R$}^{3}$ 
due to the B-field background. First of all, we assume the only 
one component of the fluctuation $A_{13}$ to be different from zero. 
The component $A_{13}$ corresponds to the $1^{--}$ glueball. 

We assume that the antisymmetric tensor field $A_{13}$ is of the form $A_{13} = \psi(u) e^{ik \cdot x}$. 
Using the metric (2.1) one obtain the differential equation for $A_{13}$: 
\begin{eqnarray}
\partial_{u} \left[\, u\,\partial_{u}\psi(u) \,\right] 
+\left\{ \frac{M^{2}}{1-\beta^{2}} \frac{1}{u^{3}} 
\left( 1-\beta^{2}(1+a^{4}u^{4}) \right) \right\}
\psi(u) = 0 \;,
\end{eqnarray}
%
In deriving the wave equation (3.2), we have chosen a particular momentum 
$k^{\mu}=(\frac{M}{\sqrt{1-\beta^2}}, 0, \frac{\beta M}{\sqrt{1-\beta^2}}, 0)$. 
Changing the variables to $z=2\ln (au)$, we have the Schr\"odinger 
type equation as 
\begin{eqnarray}
\psi'' + V^{1^{--}}_{13} \psi = 0 \;.
\end{eqnarray}
%
Here $V^{1^{--}}_{13}$ denotes the potential 
\begin{eqnarray}
V^{1^{--}}_{13}
= -\frac{1}{4} M^2 a^2 \gamma \, e^{-z} (e^{z}- a^2 \eta_{+})
(e^{z}- a^2 \eta_{-}) \;,
\end{eqnarray}
%
where 
\begin{eqnarray}
\eta_{\pm} \equiv \pm \frac{1}{a^{2}\sqrt{\gamma}} \;,
\end{eqnarray}
%
with $\gamma \equiv \frac{\beta^2}{1-\beta^2}$. The WKB approximation for this 
potential can be rewritten by using the variable $y=a^{-2}e^{z}$: 
\begin{eqnarray}
\left( n + \frac{1}{2} \right)\pi = 
\frac{1}{2} M a^2 \sqrt{\gamma} \,
\int^{\eta_{+}}_{0} dy 
\sqrt{\frac{(\eta_{+}-y)(y-\eta_{-})}{y}}  \;, 
\end{eqnarray}
%
where $n$ denotes the integer. Carrying out of this integration, we obtain 
the mass spectrum for $1^{--}$ glueball: 
\begin{eqnarray}
M^{1^{--}}_{13} = \frac{1}{16a} \, \Gamma(1/4)^{2} 
\sqrt[4]{\frac{(2n+1)^{4}}{\pi^{2}\gamma}} \;. 
\end{eqnarray}
%
As expected, the glueball masses are proportional to the inverse of the 
noncommutativity parameter. 

In the next place, we assume the only one component of the 
fluctuation $A_{03}$ to be different from zero. The component $A_{03}$ 
corresponds to the $1^{-+}$ glueball. 

The wave equation for $A_{03}$ is given by 
\begin{eqnarray}
\partial_{u} \left[\, u\,\partial_{u}\psi(u) \,\right] 
- \left\{ \frac{M^{2} \beta^{2}}{1-\beta^{2}} \frac{1+a^{4}u^{4}}{u^{3}}
 \right\}
\psi(u) = 0 \;.
\end{eqnarray}
%
In deriving the wave equation (3.8), we have set the dependencies 
$A_{03} = \psi(u) e^{ik \cdot x}$ and chosen a particular momentum 
$k^{\mu}=(\frac{M}{\sqrt{1-\beta^2}}, 0, \frac{\beta M}{\sqrt{1-\beta^2}}, 0)$. 
Under the change of variables to $z=2\ln(au)$, one obtains the Schr\"odinger 
form of the equation: 
\begin{eqnarray}
\psi'' + V^{1^{-+}}_{03} \psi = 0 \;.
\end{eqnarray}
%
Here $V^{1^{-+}}_{03}$ denotes the potential 
\begin{eqnarray}
V^{1^{-+}}_{03} 
= -\frac{1}{4} M^2 a^2 \gamma e^{-z} (1+e^{2z}) \;,
\end{eqnarray}
%
where $\gamma \equiv \frac{\beta^2}{1-\beta^2}$. Since the potential 
$V^{1^{-+}}_{03}$ takes negative value for all region of $z$, there is no 
turning point. Hence the WKB approximation for this potential 
$V^{1^{-+}}_{03}$ cannot lead the discrete mass spectrum for 
$1^{-+}$ glueball. 

The remaining components which we should investigate are $A_{01}$ and 
$A_{23}$. The components $A_{01}$ and $A_{23}$ correspond to 
the $1^{-+}$ and $1^{--}$ glueball, respectively. 
We assume that the antisymmetric tensor field $A_{\mu\nu}$ are of the form 
$A_{\mu\nu} = f(u) e^{ik \cdot x}$ and choose a particular momentum 
$k^{\mu}=(\frac{M}{\sqrt{1-\beta^2}}, 0, 
\frac{\beta M}{\sqrt{1-\beta^2}}, 0)$. For a suitable redefinition of the 
function $f(u)$ and the change of variable, we obtain 
the Schr\"odinger type equations for $A_{01}$ and $A_{23}$. The corresponding 
potentials with the Schr\"odinger equations for $A_{01}$ and $A_{23}$ are 
given as follows, 
\begin{eqnarray}
& & V^{1^{-+}}_{01} = -\frac{1}{4} M^{2}a^{2} \gamma  \, e^{-z} (1+e^{2z}) 
- \frac{e^{2z}}{1+e^{2z}} 
\left( 2-\frac{3e^{2z}}{1+e^{2z}} \right) \;, \\
& & \nonumber \\
& & V^{1^{--}}_{23} = \frac{1}{4} M^{2}a^{2} (1+\gamma)\, e^{-z} 
- \frac{2e^{2z}}{1+e^{2z}} 
+ \frac{e^{4z}}{(1+e^{2z})^{2}} \;, 
\end{eqnarray}
%
respectively. These potentials have two turning points for a certain 
region in $\gamma$. Hence the WKB approximation for the potentials 
$V^{1^{-+}}_{01}$ and $V^{1^{--}}_{23}$ implies the discrete mass spectrum 
for $1^{-+}$ and $1^{--}$ glueball. For evaluation of these mass spectra,  
numerical approach is more useful than the WKB approximation. 
More detailed analysis will be shown in \cite{nak_suz_tak}. 

The ratios of the $1^{--}$ glueball masses 
obtained by solving the wave equation for antisymmetric tensor field 
in the WKB approximation are listed in Table 4. The supergravity results 
in finite temperature Yang--Mills theory in four dimensions are also listed 
in the same table. 

\begin{table}[htbp]
\begin{center}
\begin{tabular}{|c|c|c|}
\hline 
state & $NCYM_{4}$\,(WKB) & $QCD_{4}$\,(WKB) \cite{cac_her} \\
\hline 
$1^{--}$ & 1\,(input) & 1\,(input) \\
$1^{--*}$ & 1.67 & 1.75 \\
$1^{--**}$ & 2.33 & - \\
$1^{--***}$ & 3.00 & - \\
\hline
\end{tabular}
\caption{Masses of the $1^{--}$ glueball in NCYM${}_{4}$.}
\end{center}
\end{table} 

From Table 4, we find that the difference between the supergravity 
results in the QCD and the supergravity ones in the NCYM theory is not so 
large.

%
%
\section{Conclusions and Discussions}
\setcounter{equation}{0}

In this paper, we have evaluated the ratios of the glueball masses in large $N$ 
NCSYM theories via the dual supergravity 
description. The mass spectrum of the scalar glueball $0^{++}$ and vector 
glueballs $1^{--}$ in noncommutative gauge theories are 
evaluated by solving the wave equations in the dual supergravity background.  

In evaluating the mass eigenvalues, we have 
applied the WKB analysis to the supergravity wave equations. The WKB analysis 
exhibits that the mass spectrum for the glueball is discrete with a finite 
gap due to the space-space noncommutativity. The glueball masses in 
noncommutative gauge theories depend on the noncommutativity parameter $a$ 
with dimension of length. The ratio of the glueball masses, however, 
does not depend on the noncommutativity parameter. These ratios are not 
so different from the non-supersymmetric model of QCD data. 
The supergravity dual of the noncommutative super Yang--Mills theory in a 
constant self-dual gauge field background is constructed by a certain limit of 
superposition of D3-brane and D(-1)-brane backgrounds. The effects of the 
constant self-dual gauge field background in noncommutative gauge theory can 
be estimated using the dual supergravity description. The constant self-dual 
gauge field background makes unstable glueball in noncommutative gauge theory 
stable with large D-instanton density $q$. 

The noncommutative gauge theories is not conformal due to the 
noncommutativity of space \cite{Cai_Ohta1} and the space-space noncommutativity 
is reflected in some physical quantities in the noncommutative gauge theories. 
For instance, the Wilson loops in noncommutative gauge theory exhibit area 
law behavior for the noncommutativity effects are large. The string tension, 
which can be read off from the area law, is controlled by the noncommutativity 
parameter. Similarly, the discrete mass spectra of the glueball in 
noncommutative gauge theory are caused by the space-space noncommutativity. 
The glueball masses are also controlled by the noncommutativity parameter.

The glueball mass spectrum for ordinary ${\cal N}=1$ super Yang--Mills theory 
within the Maldacena--N\`u\~nez solution has been investigated and it has been 
shown that a discrete spectrum and a mass gap for glueball can be produced 
without any sort of cut-off \cite{caceres_nunez}. It would be interesting 
subject to investigate the glueball masses in the noncommutative gauge 
theories with less than maximal supersymmetry from the noncommutative 
deformation of the Maldacena--N\`u\~nez solution. We hope to discuss this 
subject in the future.


\section*{Acknowledgments}

We would like to thank A. Sugamoto for careful reading of the manuscript and 
useful comments. One of the authors (T. N.) was partially supported by 
a Grant-in-Aid for Encouragement of Scientists (16914013) from 
the Japan Society for the Promotion of Science (JSPS).

\clearpage


\end{document}